\documentclass[10pt,a4,twocolumn]{article}

\usepackage{graphics}
\usepackage{amsmath}

\newcommand{\tr}{\mathrm{tr}\, }

\title{Dynamics of a deformable self-propelled particle \\
under external forcing}

\author{Mitsusuke Tarama\thanks{ Department of Physics, Kyoto University, Kyoto, 606-8502, Japan. } ~and Takao Ohta\thanks{ Department of Physics, Kyoto University, Kyoto, 606-8502, Japan. E-mail adress: takao@scphys.kyoto-u.ac.jp } }

\begin{document}

\maketitle

\begin{abstract}
We investigate dynamics of a self-propelled deformable particle under external field in two dimensions based on the model equations for the center of mass and a tensor variable characterizing deformations.  We consider two kinds of external force. One is a gravitational-like force which enters additively in the time-evolution equation for the center of mass. The other is an electric-like force supposing that a dipole moment is induced in the particle. This force is added to the equation for the deformation tensor.  It is shown that a rich variety of dynamics appears by changing the strength of the forces and the migration velocity of self-propelled particle.
\end{abstract}

\section{Introduction}\label{sec:intro}
Dynamics of self-propelled objects have attracted much attention recently from the view point of various new orders far from equilibrium. In order to elucidate the new dynamical states, one needs to develop the theory of nonlinear dynamics and  non-equilibrium statistical physics. There are many kinds of self-propelled motions both in biological and non-biological systems. Migration of microorganisms such as living cells and bacteria
\cite{Keren,Bosgraaf,Li,Maeda,Wada,Ishikawa} are typical examples in biological systems.   Artificial models have been introduced for self-organized microswimmers \cite{Kruse,Fu,Yeomans}. On the other hand, there are investigations  of self-propulsion in non-biological systems such as  oily droplets in surfactant solution  \cite{Nagai,Toyota1}, colloid composites under external field or with chemical reactions  \cite{Sagues,Kapral1,Kapral2}  and  Janus particles 
\cite{Sano,Squires,Showalter}. Marangoni effects due to chemical reactions and electrophoresis are the origins of the self-propulsion of the above systems. See a review article \cite{Ebbens} for the related nano-systems.  It is known that theoretical formulation of self-propulsion is non-trivial even for a single particle.

 In a previous paper, one of the present authors (TO) and Ohkuma introduced a set of model equations for a deformable self-propelled particle \cite{OhtaOhkuma}.
 The model equations have been constructed by symmetry argument and hence from a general view point of a dynamical system.  After that, those equations  together with higher modes of deformations have been derived by the singular perturbation method
for an excitable reaction-diffusion system \cite{Krischer} both in two dimensions \cite{OOS} and in three dimensions \cite{SHO}. Various dynamics of motions are obtained  by changing the migration velocity and the softness of the particle \cite{OhtaOhkuma,Hiraiwa1,Hiraiwa2}. The collective dynamics of the particles with orientational interactions have been investigated in two dimensions \cite{OhtaOhkuma,OhkumaOhtaFull,Itino}.  
It is  mentioned that deformable self-propulsion was studied by Shapere and Wiczek to investigate efficiency of swimming motion \cite{Wilczek}. It is also mentioned that  there are a number of theoretical studies of collective dynamics of {\it non}-deformable self-propelled particles. See Ref.~\cite{Ramaswamy} and the earlier papers cited therein.

In the present paper, we study dynamics of a deformable self-propelled particle under external force in two dimensions. The system we consider is a single isolated particle but has internal degrees of  freedom due to deformability. It should be noted that, even when the external forcing is absent, there is a bifurcation between a straight motion and a circular motion along a closed trajectory \cite{OhtaOhkuma}. Therefore, by adding an external force, there occurs a conflict or frustration between the circular motion and the forced straight motion. As a result,  non-trivial behavior of motion and bifurcations are exhibited by changing the magnitude of the external force.  A model of self-propulsion which undergoes a circular motion has also been introduced in Ref.~\cite{Teeffelen}. 

We consider two kinds of external forcing. One is a gravitational-like force which enter additively in the equation of motion for the center of mass. A charged particle under electric field is also governed by this euation. The other is an electric-like force which is supposed to produce an electric dipole in the particle. The force is added to the equation for the deformation tensor. We shall carry out numerical simulations of the set of the time-evolution equations to obtain a dynamical phase diagram. Analytical study will also be developed to reproduce some of the motions and the bifurcations.

The organization of this paper is as follows. In the next section (section \ref{sec:model}), we introduce the model equation. Numerical results for the gravitational-like forcing is given in section \ref{sec:numerical_results_1}. We will show that a circular-drift motion occurs in this case, where a particle causes a drift motion to the direction perpendicular to the external force.  This motion is analyzed in detail in section \ref{sec:circular_drift}. Numerical simulations for the electric-like forcing are described in section \ref{sec:numerical_results_2}.  Analytical study of the gravitational-like force is presented in section \ref{sec:analysis_bifurcations_1} whereas that of the electric-like force is given in section \ref{sec:analysis_bifurcations_2}. Summary and discussion are given in section \ref{sec:discussion}.

\section{Model equations}\label{sec:model}

We consider the following set of equations of motion in two dimensions for the velocity of the center of mass ${\bf
v}=(v_1, v_2)$ and the symmetric tensor $S_{\alpha \beta}$ characterising deformations 
\begin{gather}
\frac{d v_{\alpha}}{d t} = \gamma v_{\alpha} - |v^2| v_{\alpha} - a S_{\alpha \beta} v_{\beta} + g_{\alpha} 
 \label{eq:1.1} \\
\frac{d S_{\alpha \beta}}{d t} = - \kappa S_{\alpha \beta} +b \left( v_{\alpha} v_{\beta} -\frac{1}{2}|v^2| \delta_{\alpha \beta} \right) 
+ Q_{\alpha \beta} , 
 \label{eq:1.2}
\end{gather}
where 
\begin{gather}
Q_{\alpha \beta}  =h \left( E_{\alpha} E_{\beta} - \frac{\left| E^2 \right|}{2}  \delta_{\alpha \beta} \right). 
 \label{eq:Q}
\end{gather}
This type of external force has been considered in a phase separated droplet under electric field \cite{Orihara}.
The coefficient $\gamma$ may change the sign and $\kappa$ is positive.  $a$ and $b$ are coupling constants. The tensor $S_{\alpha \beta}$ is traceless to make the area of the particle (volume of the particle in three dimensions) constant and defined by
$S_{\alpha \beta} = s \left( n_{\alpha} n_{\beta}-\frac{1}{2}\delta_{\alpha, \beta} \right)$, where  the  unit vector ${\bf n}$ is parallel to the long axis of a deformed elliptical particle and $s>0$ is the degree of deformation from a circular shape. 
The external force $g_{\alpha}$ is assumed to be  given by ${\bf g} = (0, -g)$ with $g > 0$. The other external force $E_{\alpha}$ in $Q_{\alpha \beta}$ is applied as ${\bf E} =  (1, 0)$. The left-right and the up-down symmetry of the system allow us to put $h>0$ for fixed signs of $a$ and $b$.
 The set of equations~(\ref{eq:1.1}) and (\ref{eq:1.2}) without the external forces has been introduced in Ref.~\cite{OhtaOhkuma} and has been derived from an excitable reaction-diffusion system not only in two dimensions but also in three dimensions  
\cite{OOS,SHO}.

Equations~(\ref{eq:1.1}) and (\ref{eq:1.2}) can be written as 
\begin{eqnarray}
\frac{d v}{d t} &=& \gamma v -v^3 -\frac{a}{2} s v \cos 2\psi - g \sin \phi \label{eq:1.3} \\
\frac{d \phi}{d t} &=& -\frac{a}{2} s \sin 2\psi - \frac{g}{v} \cos \phi \label{eq:1.4} \\
\frac{d s}{d t} &=& -\kappa s +b v^2 \cos 2\psi +h \cos 2\theta \label{eq:1.5} \\
\frac{d \theta}{d t} &=& -\frac{b}{2 s} v^2 \sin 2\psi -\frac{h}{2 s} \sin 2\theta , \label{eq:1.6}
\end{eqnarray}
where we have put 
\begin{eqnarray}
v_1 &=& v \cos \phi,\label{eq:v1} \\ 
v_2 &=& v \sin \phi, \label{eq:v2} \\ 
n_1 &=&\cos \theta,\label{eq:n1} \\ 
n_2 &=& \sin \theta,\label{eq:n2} 
\end{eqnarray}
with $v$ and $s$ positive values, 
and
 \begin{eqnarray} 
 \psi = \theta -\phi .
 \label{eq:psi}
 \end{eqnarray}
Putting Eqs.~(\ref{eq:1.4}) and (\ref{eq:1.6}) together, we have 
\begin{eqnarray}
\frac{d \psi}{d t} 
 &=& -\frac{1}{2} \left(  \frac{b v^2}{s}  - a s  \right) \sin 2\psi  + \frac{g}{v} \cos \phi \nonumber \\
 &-&\frac{h}{2 s} \sin 2(\phi + \psi)
. \label{eq:1.8}
\end{eqnarray}

Equations~(\ref{eq:1.1}) and  (\ref{eq:1.2}) in the absence of $g$ and $Q_{\alpha \beta}=0$  exhibit a drift bifurcation. That is, when $\gamma < 0$,  a stable state is a motionless state, and when $\gamma > 0$,  a particle undergoes a self-propelled motion. 
In the latter case, there is another bifurcation \cite{OhtaOhkuma}. The constant defined by
\begin{equation}
 B= \frac{ab}{2\kappa} 
 \label{eq:B}
\end{equation}
plays an important role. If $B$ is positive as we assume throughout this paper, a particle undergoes a  straight motion in some direction determined by the initial condition for $0 < \gamma < \gamma_c$. The bifurcation threshold $\gamma_c$ is given by \cite{OhtaOhkuma}
\begin{equation}
\gamma_c \equiv \frac{\kappa^2}{a b} +\frac{\kappa}{2} = \frac{\kappa (1+B)}{2B} . 
 \label{eq:1.7} 
\end{equation}
There are two cases of the straight motion depending on the sign of $b$. If $b$ is positive, the particle elongates along the direction of the migration velocity whereas, if it is negative, the elongation is perpendicular to the velocity. 
This straight motion loses its stability  for $\gamma > \gamma_c$ and a periodic motion along a closed circle appears to be stable \cite{OhtaOhkuma}. 

In the following sections, we study the interplay between the circular motion and the straight motion forced by the external fields.

\section{Numerical Results I}\label{sec:numerical_results_1}

\begin{figure}[t]
  \begin{center}
  \resizebox{0.35\textwidth}{!}{
  \includegraphics{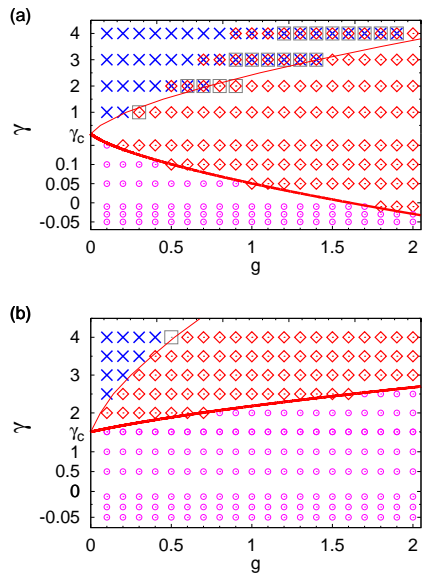}
}
    \caption{(Colour on-line)
      Phase diagram on the $\gamma-g$ plane for (a) $\kappa = 0.2$ and (b) $\kappa = 0.75$. 
      The symbols indicate the following motions; the circular-drift motion (cross), the zigzag-1 motion (diamond), the zigzag-2 motion (square), and the straight-falling motion (circle). Note that there are coexistence regions. 
      The thin solid line is the saddle homoclinic-orbit bifurcation boundary determined numerically from  the reduced equations (\ref{eq:3-1.1}) and (\ref{eq:3-1.2}). 
      The thick solid line is the Hopf bifurcation boundary obtained by Eq.~(\ref{eq:N1.5}). 
          } \label{fig:diagram_1}
  \end{center}
\end{figure}

In this section, we show the results of numerical simulations of the model equations (\ref{eq:1.3}) - (\ref{eq:1.6}) for the gravitational-like external force, i.e., 
with $Q_{\alpha\beta}=0$.  
The fourth-order Runge-Kutta method is employed with time increment $\delta t=10^{-4}$. The coupling coefficients $a$ and $b$ are fixed as $a = -1.0$ and $b = -0.5$. 
We have obtained the phase diagram on the $\gamma$-$g$ plane for $\kappa = 0.2$ and $\kappa = 0.75$, as shown in Fig.~\ref{fig:diagram_1}.  
The bifurcation threshold is given by $\gamma_c=0.18$ for $\kappa=0.2$ and $\gamma_c=1.5$ for $\kappa=0.75$.  There are four different motions: 
 a circular-drift motion,  a zigzag-1 motion,  a zigzag-2 motion, and a straight-falling motion as we shall explain below.

\begin{figure}[t]
  \begin{center}
   \resizebox{0.4\textwidth}{!}{
  \includegraphics{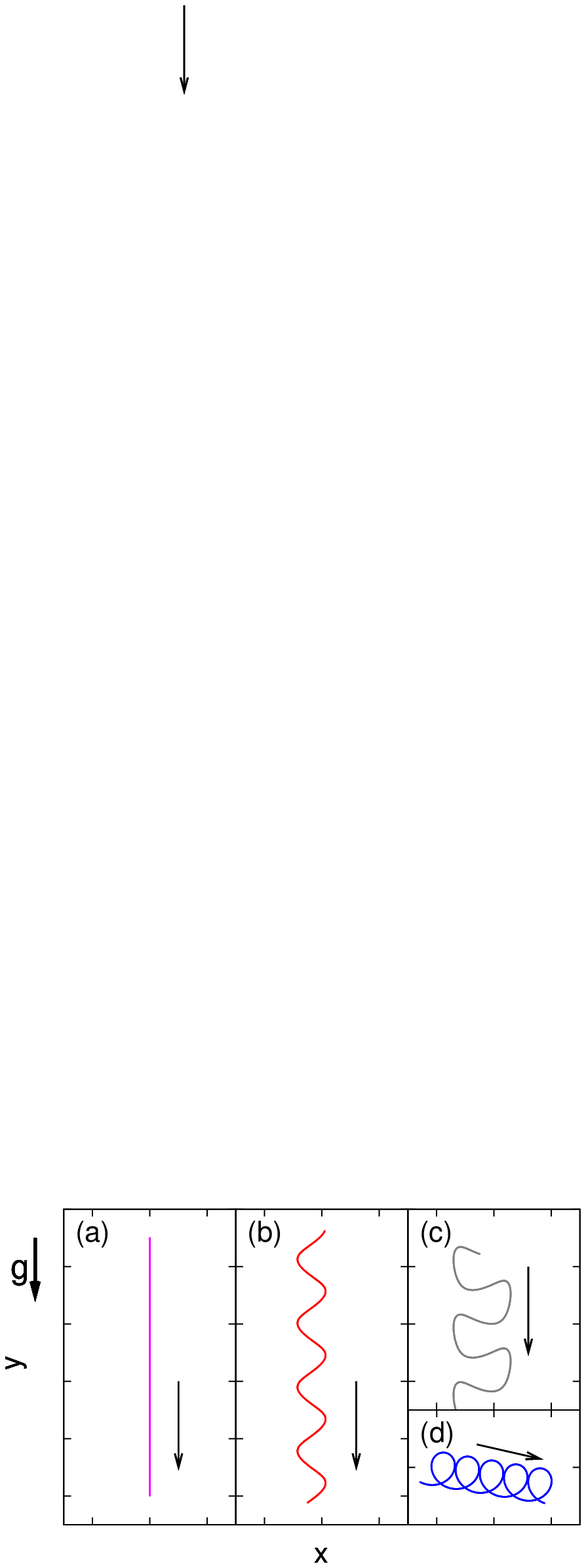}
}
    \caption{(Colour on-line)
      Trajectory of (a) the straight-falling motion, (b) the zigzag-1 motion, (c) the zigzag-2 motion and (d)  the circular-drift motion. The arrows indicate the direction of migration.      The parameters are chosen as $\gamma = 2$ and $\kappa = 0.75$ for (a)  and $\gamma = 3$ and $\kappa = 0.2$ for (b), (c) and (d). 
    } \label{fig:original_traces_1}
  \end{center}
\end{figure}

First of all, we consider the region $\gamma < 0$, where a particle is motionless when the external force is absent. 
When the external force ${\bf g} = (0, -g)$ is added and the magnitude is small, it causes a straight motion in the direction of the external force as expected.  
Since $b$ is chosen to be negative, the elongation is perpendicular to the external force. 
We call this trivial motion a straight-falling motion. It should be emphasized, however, that a non-trivial behavior occurs for large magnitudes of $g$ as can be seen around $g \approx 2$ and $\gamma \approx -0.01$ in Fig.~\ref{fig:diagram_1}(a) where the straight motion is unstable and the zigzag-1 motion appears. The straight-falling motion and the zigzag-1 motion are displayed in Fig.~\ref{fig:original_traces_1}(a) and Fig.~\ref{fig:original_traces_1}(b), respectively. The bifurcation between the straight-falling motion and the zigzag-1 motion exists also for $0 < \gamma < \gamma_c$ as shown in Fig.~\ref{fig:diagram_1}(a).  When the external force is absent, this is the region that the particle undergoes a straight self-propelled motion, whose direction depends on the initial conditions. Figure~\ref{fig:diagram_1}(b) for $\kappa=0.75$ indicates that  such a bifurcation does not exist for $\gamma < \gamma_c$. These results can be understood theoretically as described in section \ref{sec:analysis_bifurcations_1}.

Next, we consider the case $\gamma > \gamma_c$ where a circular motion appears when $g=0$. 
In the region indicated by the cross in Fig.~\ref{fig:diagram_1} where the magnitude of the external  force $g$ is finite but small, a particle takes a circle-like motion. 
However, in this case, the center of the circle drifts to some direction as shown in Fig.~\ref{fig:original_traces_1}(d). We call this motion a circular-drift motion.  The drift direction asymptotically in time is determined uniquely for given values of the parameters.   But it is neither equal to the direction of the external force nor determined  by the initial condition.  
We discuss this property in detail in section \ref{sec:circular_drift}. 
For larger values of  the external  force,  a zigzag-1 motion appears.
For $\kappa = 0.2$ in Fig.~\ref{fig:diagram_1}(a), there is a coexisting parameter region of the circular-drift motion and the zigzag-1 motion. It is noted that 
there is  another zigzag-like motion,  called zigzag-2 motion, in  the coexisting region, which is shown in Fig.~\ref{fig:original_traces_1}(c). 
In the case $\kappa = 0.75$ in Fig.~\ref{fig:diagram_1}(b), the zigzag-2 motion 
has been observed only in the small region near $g=0.5$ and $\gamma=4$.
When the external  force is further increased, the zigzag-1 motion becomes unstable  and a straight-falling motion appears for $\kappa = 0.75$.  This is contrast to the case of $\kappa = 0.2$ where the straight-falling motion does not appear for $\gamma>\gamma_c$.

\begin{figure}[t]
  \begin{center}
    \resizebox{0.4\textwidth}{!}{
  \includegraphics{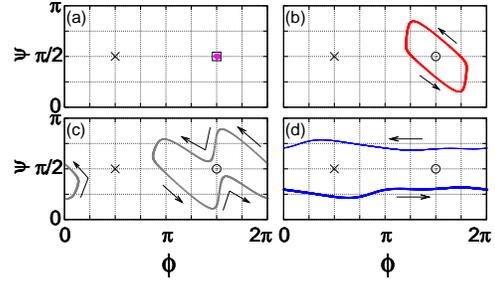}
}
    \caption{(Colour on-line)
      Attractors in the $\psi$-$\phi$ plane for (a) the straight-falling motion, (b) the zigzag-1 motion, (c) the zigzag-2 motion and (d)  the circular-drift motion. These  are  obtained by solving Eqs.~(\ref{eq:1.3}) - (\ref{eq:1.6}) numerically. The parameters for each motion are the same as those in Fig.~\ref{fig:original_traces_1}.   The arrows indicate the direction of motion. The square in (a) and the circle in (b), (c) and (d) indicate the stable and the unstable fixed points, respectively, and $\times$ indicates the saddle point. 
    } \label{fig:phipsi_1}
  \end{center}
\end{figure}

It is convenient to represent the motions asymptotically in time  in the $\phi$-$\psi$ plane where $\phi$ and $\psi$ have been defined by (\ref{eq:v1}) and (\ref{eq:v2}), and (\ref{eq:psi}) respectively. 
It is noted  that the equations (\ref{eq:1.3}) - (\ref{eq:1.6}) are invariant under the transformations $\phi \to \phi+ 2 \pi$ and $\psi \to \psi + \pi$. 
Therefore, we may consider only the restricted range $0< \phi< 2\pi$ and $0< \psi< \pi$.
The straight-falling motion  is represented by the stable fixed point $(\phi, \psi) = (3\pi/2, \pi/2)$, as shown in Fig.~\ref{fig:phipsi_1}(a). 
The zigzag-1 and -2 motions are limit-cycles around this fixed point  as shown in Fig.~\ref{fig:phipsi_1}(b) and (c) respectively. 
We discuss the difference of these motions in detail in the section \ref{sec:analysis_bifurcations_1}. 
The attractor of the circular-drift motion is shown in  Fig.~\ref{fig:phipsi_1}(d).

\begin{figure}[t]
  \begin{center}
    \resizebox{0.4\textwidth}{!}{
  \includegraphics{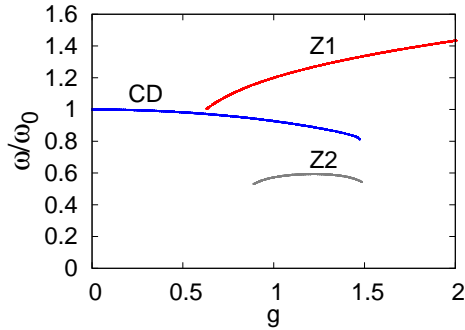}
}
    \caption{(Colour on-line)
      The normalized frequency $\omega/\omega_0$ of the circular-drift motion (CD), the zigzag-1 motion (Z1), and the zigzag-2 motion (Z2) as a function of $g$ for $\gamma = 3$ and $\kappa = 0.2$. 
    } \label{fig:frequency_1}
  \end{center}
\end{figure}

Now we discuss the coexistence region  of the circular-drift motion, the zigzag-1 motion and the zigzag-2 motion in Fig.~\ref{fig:diagram_1}(a). 
Since the trajectories of these motions in Fig.~\ref{fig:phipsi_1} are periodic  in the $\phi$-$\psi$ plane, one can define the period $T$ for each motion. 
In Fig.~\ref{fig:frequency_1}, we display the frequency of these three motions in the coexisting region for $\gamma=3$ and $\kappa=0.2$, which is normalized by the frequency $\omega_0$ of the circular motion  without the external forcing. 
We find that the frequency of the zigzag-1 motion becomes close to that of the circular-drift motion in the vicinity  where the zigzag-1 motion becomes unstable. 
The frequency of the zigzag-2 motion is approximately a half of that of the zigzag-1 motion.

\section{Circular-Drift Motion} \label{sec:circular_drift}

\begin{figure}[t]
  \begin{center}
    \resizebox{0.4\textwidth}{!}{
  \includegraphics{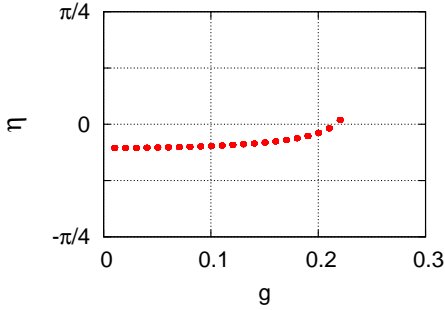}
}
   \caption{(Colour on-line)
      Angle of the mean displacement of the circular-drift motion as a function of the external force $g$
     for $\gamma = 3$ and  $\kappa = 0.75$. 
    } \label{fig:meanangle_0.75}
  \end{center}
\end{figure}

One of the most interesting properties of the circular-drift motion is, as shown in Fig.~\ref{fig:original_traces_1}(d), 
 that the direction of the drift motion is almost perpendicular to the external force ${\bf g} = (0, -g)$ with $g>0$. 
The drift direction is defined quantitatively as follows. It should be noted that although the trajectory of a circular-drift motion in the real space is not periodic, it is precisely periodic in the velocity space. Therefore, we may define the position after one period starting from the coordinate origin as  
\begin{eqnarray}
\left[
\begin{array}{c}
\bar{X} \\ \bar{Y} 
\end{array}
\right]
&=& 
 \int_{0}^{T} dt\  v(t) 
 \left[
 \begin{array}{c}
 \cos \phi(t) \\ \sin \phi(t)
 \end{array}
 \right]
 \notag \\ 
&=&
 \int_{0}^{2\pi} d \phi \left| \frac{d \phi}{d t} \right|^{-1}  v( \phi )
 \left[
 \begin{array}{c}
 \cos \phi  \\ \sin \phi
 \end{array}
 \right]   ,  
 \ \ \ \ \ \ \ \ 
\label{eq:A2.1}
\end{eqnarray}
where $T$ is the period. 
The angle of the mean displacement with respect to the x-axis is given through the relation
\begin{equation}
\tan \eta =   \frac{\bar{Y}}{\bar{X} } . 
 \label{eq:A2.2} 
\end{equation}
In Fig.~\ref{fig:meanangle_0.75}, we show the angle $\eta$ as a function of $g$ for $\gamma=3$ and $\kappa=0.75$. Since the system possesses the right-left symmetry, one may restrict to $-\pi/2<\eta <\pi/2$ without loss of generality. The angle $\eta$ is  slightly smaller than $0$, i.e., $\eta\approx-0.05\pi$ and gradually increases by increasing $g$ up to the bifurcation threshold $g \approx 0.215$. One unexpected phenomenon is that the value of $\eta$ becomes positive in the very vicinity of the threshold although the magnitude is extremely small. This means that the particle gradually moves upward on an average while drifting  to the right. Since the particle migrates by consuming the internal energy, this does not violate energy conservation. However, we do not have any definite explanation of this phenomenon which occurs in an extremely restricted parameter region.

\section{Numerical Results II}\label{sec:numerical_results_2}

In this section, we show the numerical results of self-propulsion in  the electric-like external field. 
We have solved Eqs.~(\ref{eq:1.3}) - (\ref{eq:1.6}) 
with $g_{\alpha}=0$ and $Q_{\alpha \beta} \ne 0$ by using the fourth Runge-Kutta method with the time increment $\delta t=10^{-4}$ for  $\kappa=0.5$,  $a = -1.0$ and $b = -0.5$. Therefore $B\equiv a b /2\kappa=0.5>0$.

\begin{figure}[t]
	\begin{center}
	\resizebox{0.45\textwidth}{!}{
	  \includegraphics{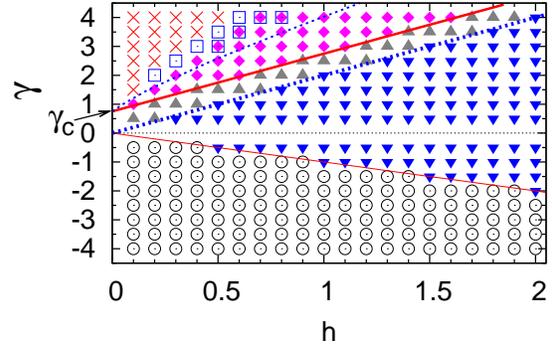}
	}
\caption{(Colour on-line)
      Phase diagram on the $\gamma$-$h$ plane for $\kappa = 0.5$ obtained by numerical simulations of Eqs.~(\ref{eq:1.3}) - (\ref{eq:1.6}). The symbols indicate the circular motion (cross), the zigzag-1 motion (diamond), the zigzag-2 motion (square), and the motionless state (circle). 
      The up and down triangles indicate the straight motion with $\bf{n}$ perpendicular and parallel to $\bf{E}$ respectively.  The transition boundary between them 
       is consistent with the analytical results given by Eq.~(\ref{eq:3-2.100}), which is shown by the thick dotted line. 
      The thick solid line is the Hopf bifurcation boundary given by $h=h_{H}$ 
      whereas  the thin solid line is the pitchfork bifurcation boundary given by Eq.~(\ref{eq:3-2.19}). 
      The thin dotted line is the bifurcation boundary between the circular motion and the zigzag-1 motion 
      obtained numerically from the reduced equations (\ref{eq:3-2.1}) and (\ref{eq:3-2.2}).  
          } \label{fig:diagram_2}
	\end{center}
\end{figure}

\begin{figure}[t]
  \begin{center}
	\resizebox{0.45\textwidth}{!}{
	  \includegraphics{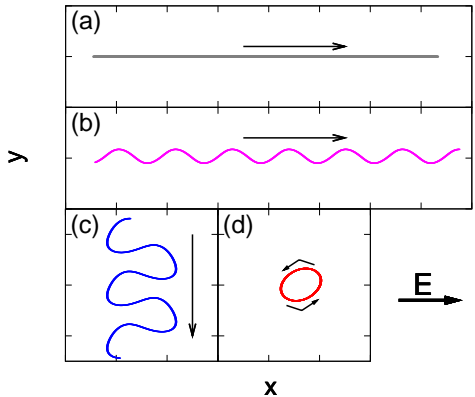}
}
    \caption{(Colour on-line)
    Trajectories of the center of mass in the real space for $\gamma = 3$ (a) a  straight motion for $h=1.3$, (b)  a zigzag-1 motion for $h=0.6$, (c) a zigzag-2 motion for $h=0.5$, and (d) a counter-clockwise circular motion for $h=0.3$  obtained by solving Eqs.~(\ref{eq:1.3}) - (\ref{eq:1.6}) numerically. 
     } \label{fig:original_traces_2}
  \end{center}
\end{figure}

\begin{figure}[t]
  \begin{center}
\resizebox{0.45\textwidth}{!}{
	  \includegraphics{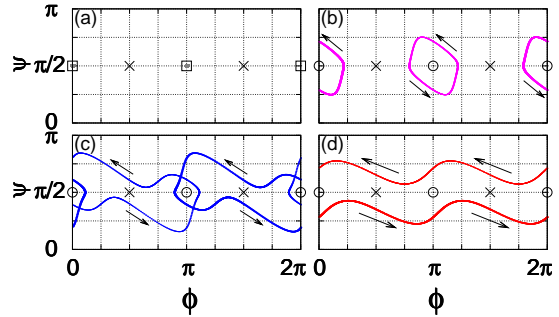}
}
    \caption{(Colour on-line)
     Trajectories on the $\psi$-$\phi$ plane for (a) the straight motion, (b) the zigzag-1 motion, (c) the zigzag-2 motion, and (d) the circular motion, obtained by solving Eqs.~(\ref{eq:1.3}) - (\ref{eq:1.6}) numerically. 
      The white circles and the white squares indicate the unstable and the stable fixed points, and the crosses indicate a saddle point. 
      The values of the parameters are the same as those of Fig.~\ref{fig:original_traces_2}.
    } \label{fig:phipsi_2}
  \end{center}
\end{figure}

Figure~\ref{fig:diagram_2} displays, on the $\gamma$-$h$ space, the phase diagram  of a variety of dynamical states: 
 a circular motion, a zigzag-1 motion, a zigzag-2 motion, and a straight motion. 
Figure~\ref{fig:original_traces_2} shows the trajectory of the center of mass of each motion in the real space. 
We also show the trajectories of these motions in the $\phi$-$\psi$ plane in Fig.~\ref{fig:phipsi_2}. 
When the magnitude of the external force $h$ is large enough, a particle undergoes a straight motion 
for all $\gamma$. This occurs in the region indicated by the up and down triangles in Fig.~\ref{fig:diagram_2}.
In this straight motion, even though we have chosen $b<0$, the direction of the deformation is parallel to the external force in the region of the down triangles. 
The elongation becomes perpendicular for smaller values of $h$ as indicated by the top triangles in Fig.~\ref{fig:diagram_2}. 
When the magnitude of the force $h$ is decreased, the motion of a particle is affected by the three stable states which occur when the external force is absent \cite{OhtaOhkuma}: the motionless state for $\gamma<0$, the straight motion for $0<\gamma<\gamma_c$, and the circular motion for $\gamma>\gamma_c$. 
For $\gamma<0$, a particle does not move  under a finite but weak external force, and therefore, the stable state is a motionless state. 
By increasing the magnitude $h$, a bifurcation occurs from the motionless state to a straight motion as shown in Fig.~\ref{fig:diagram_2}.
For $0<\gamma<\gamma_c$, a particle undergoes a straight motion irrespective of the values of the external force. 
When $\gamma>\gamma_c$, a particle undergoes an circular motion along an elliptically-deformed trajectory as in Fig.~\ref{fig:original_traces_2}(d) under a finite but weak external force in the region indicated by the crosses in Fig.~\ref{fig:diagram_2}.
Between this circular motion and the straight motion, there is a region indicated by the diamonds  in Fig.~\ref{fig:diagram_2} where a zigzag-1 motion occurs as displayed in Fig.~\ref{fig:original_traces_2}(b). 
When $\gamma-\gamma_c>0$ is large enough, there appears another motion, which is called a zigzag-2 motion  between the circular motion and the zigzag-1 motion as indicated by the square in Fig.~\ref{fig:diagram_2}. The trajectory is displayed in  Fig.~\ref{fig:original_traces_2}(c).

\begin{figure}[t]
  \begin{center}
	\resizebox{0.45\textwidth}{!}{
	  \includegraphics{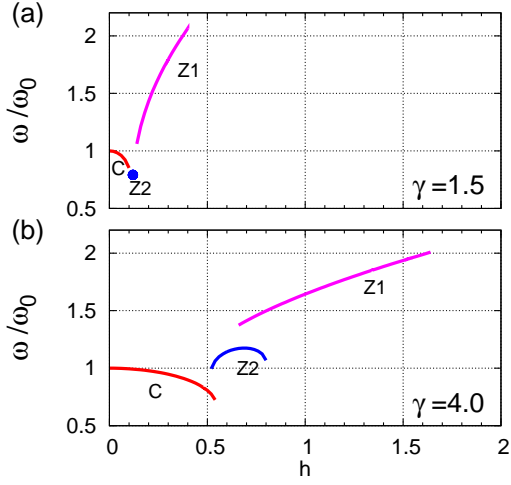}
	}
    \caption{(Colour on-line)
      Normalized frequency of the circular motion (C), the zigzag-1 motion (Z1) and the zigzag-2 motion (Z2) for (a) $\gamma=1.5$ and (b) $\gamma=4$. 
    } \label{fig:frequency_2}
  \end{center}
\end{figure}

 It is evident in Fig.~\ref{fig:diagram_2}  that there is a region where the zigzag-2 motion and the zigzag-1 motion coexist. 
Although not clearly shown, there is a small region $0.52< h <0.54$ for $\gamma=4$ where the circular motion and the zigzag-2 motion coexist. These are also seen by analyzing the frequency of the motions. That is, the attractors of the circular motion and the zigzag-1 motion and the zigzag-2 motion are periodic in the $\phi$-$\psi$ space as shown in Fig.~\ref{fig:phipsi_2}. 
Therefore, one can define the frequency for these three motions, which is depicted in Fig.~\ref{fig:frequency_2} where the frequency is normalized by that of the no external force  $\omega_0$ for (a)  $\gamma=1.5$ and (b) $\gamma=4$. Figure \ref{fig:frequency_2}(b) clearly indicates that there are coexistence regions of the circular motion and the zigzag-2 motion, and the zigzag-2 motion and the zigzag-1 motion.

\section{Analysis of the bifurcations I}\label{sec:analysis_bifurcations_1}

The set of time-evolution equations (\ref{eq:1.3}) - (\ref{eq:1.6}) with $Q_{\alpha\beta}=0$  is complicated for  theoretical analysis.
Here we make a reduction of the variables by putting $d v /d t =d s / d t=0$. This approximation is justified 
if the variables $s$ and $v$ relax rapidly compared with $\phi$ and $\psi$. Unfortunately, however, this is not generally guaranteed in the present problem. Therefore, we need to check the results of the reduced system by comparing with the numerical results of the original system governed by Eqs.~(\ref{eq:1.3}) - (\ref{eq:1.6}).

Eliminating $s$ and $v$ from Eqs.~(\ref{eq:1.3}) - (\ref{eq:1.6}),
 one obtains
\begin{gather}
\frac{d \phi}{d t} =  F\left( \phi, \psi \right)
 \label{eq:3-1.1} \\
\frac{d \psi}{d t} =  G\left( \phi, \psi \right),  
 \label{eq:3-1.2}
\end{gather}
where 
\begin{gather}
F\left( \phi, \psi \right)  =  -B v^2 \cos 2\psi \sin 2\psi -\frac{g}{v} \cos \phi 
 \label{eq:3-1.3}  \\
G\left( \phi, \psi \right)  =  -\frac{\kappa}{2} \tan 2\psi -F\left( \phi, \psi \right) . 
 \label{eq:3-1.4}
\end{gather}
and $B$ has been defined by Eq.~(\ref{eq:B}).
The magnitude $s$ is  given from  Eq.~(\ref{eq:1.5}) by 
\begin{equation}
s\left( \phi, \psi \right)  = \frac{b}{\kappa} v^2 \cos 2\psi . 
 \label{eq:3-1.5}
\end{equation}
In the same way,  the magnitude of the velocity for small value of $g$ is determined from Eq.~(\ref{eq:1.3}), as the largest positive root of the following cubic equation, 
\begin{equation}
v^3 -3 p v +2 q = 0  , 
 \label{eq:3-1.21}
\end{equation}
where 
\begin{eqnarray}
p(\phi, \psi) &=&\frac{\gamma}{3 \left( 1 +B \cos^2 2\psi \right)} 
 \label{eq:3-1.8} \\
q(\phi, \psi) &=& \frac{g \sin \phi }{2 \left( 1 +B\cos^2 2\psi \right) } . 
 \label{eq:3-1.9}
\end{eqnarray}
We have verified, by numerical simulations of the reduced equations (\ref{eq:3-1.1}) and (\ref{eq:3-1.2}), that all types of the motions except for the zigzag-2 motion are reproduced for finite values of  $g$. The dynamical phase diagram is quite similar to Fig.~\ref{fig:diagram_1} obtained from the original equations (\ref{eq:1.3}) - (\ref{eq:1.6}).

If  the variable $s$ is also retained as a slow variable, one obtains 
\begin{eqnarray}
\frac{d \phi}{d t} &=& - \frac{a}{2} s \sin 2\psi - \frac{g}{v} \cos \phi
 \label{eq:A1.1} \\
\frac{d s}{d t} &=& -\kappa s +b v^2 \cos 2\psi
 \label{eq:A1.2} \\
\frac{d \psi}{d t} &=& - \frac{b v^2 -a s^2}{2 s} \sin 2\psi +\frac{g}{v} \cos \phi , 
 \label{eq:A1.3}
\end{eqnarray}
The magnitude of the velocity is given by the largest positive root of the cubic equation 
\begin{equation}
v^3 -3 \bar{p} v +2 \bar{q} = 0 ,
 \label{eq:3-1.23}
\end{equation}
where
\begin{eqnarray}
\bar{p}( \phi, \psi ) &=& \gamma/3 - (a/6) s \cos 2\psi
 \label{eq:A1.6} \\
\bar{q}( \phi, \psi ) &=& (g/2) \sin \phi  . 
 \label{eq:A1.7}
\end{eqnarray}
It is verified numerically that these three-variable equations exhibit the zigzag-2 motion as well as other three kinds of motions mentioned above. However, since the analysis of the  three-variable system is complicated, we carry out the stability analysis for the two-variable equations (\ref{eq:3-1.1}) and (\ref{eq:3-1.2}) below.

Equations (\ref{eq:3-1.1}) and (\ref{eq:3-1.2})  have
 two fixed points, $(\phi, \psi)=(\pi /2, \pi/2)$ and $(3 \pi /2, \pi/2)$. 
 We define the stability matrix as
 \begin{equation}
L\left( \phi, \psi \right)  =
\left[
\begin{array}{cc}
\partial_{\phi}  F(\phi, \psi) &~  \partial_{\psi}  F(\phi, \psi) \\
\partial_{\phi}  G(\phi, \psi) &~  \partial_{\psi}  G(\phi, \psi)
\end{array}
\right], 
 \label{eq:3-1.10} 
\end{equation}
where $\partial_{\phi}F=\partial F/\partial \phi$.
The stability matrix at the fixed point $( \pi/2, \pi/2 )$ is given by
\begin{equation}
L\left( \frac{\pi}{2}, \frac{\pi}{2} \right) = 
\left[
\begin{array}{cc}
g /v &~ -2 B v^2 \\
-g /v &~ -\kappa + 2 B v^2
\end{array}
\right]  , 
\label{eq:3-1.11}
\end{equation}
where $v$ is a positive root of Eq.~(\ref{eq:3-1.21}) with $p=p_0$ and $q=q_0$ where  
\begin{gather}
p_0 \equiv\frac{\gamma}{3(1+B) }
 \label{eq:3-1.12} \\
q_0 \equiv \frac{g }{2(1+B)} . 
 \label{eq:3-1.13}
\end{gather} 
The trace and the determinant of the stability matrix are calculated as
\begin{gather}
\tr L\left( \pi /2, \pi /2 \right) = g/v -\kappa + 2B v^2 
 \label{eq:3-1.14}  \\
\det L\left( \pi /2, \pi /2 \right) = -g \kappa /v  . 
 \label{eq:3-1.15} 
\end{gather}
Since $v$ should be positive, the determinant is always negative so that we find the fixed point $(\pi/2, \pi/2)$ is a saddle point. 

The stability matrix for the other fixed point $(3\pi/2, \pi/2)$ corresponding  to the straight-falling motion is given by 
\begin{equation}
L\left( \frac{3\pi}{2}, \frac{\pi}{2} \right) = 
\left[
\begin{array}{cc}
-g /v &~ -2 B v^2 \\
g /v &~ -\kappa + 2 B v^2
\end{array}
\right]  , 
 \label{eq:3-1.16}
\end{equation}
from which one obtains
\begin{gather}
\tr L\left( 3\pi /2, \pi /2 \right) = -g /v -\kappa + 2B v^2 
 \label{eq:3-1.17}\\
\det L\left( 3\pi /2, \pi /2 \right) = g \kappa /v . 
 \label{eq:3-1.18}
\end{gather}
where $v$ is a positive root of the cubic equation (\ref{eq:3-1.21}) with $p=p_0$ and $q=-q_0$. 
Since the determinant (\ref{eq:3-1.18}) is always positive, a Hopf bifurcation occurs at $\tr L(3\pi/2, \pi/2)=0$.
At this fixed point, 
Eq.~(\ref{eq:3-1.21}) can be written as
\begin{equation}
(1+B) v^3 -\gamma v -g = 0 . 
 \label{eq:N1.3}
\end{equation}
Using this equation, the derivative of the trace (\ref{eq:3-1.17}) with respect to $g$ is calculated as \begin{equation}
\frac{\partial}{\partial g}\tr L\left( \frac{3\pi}{2}, \frac{\pi}{2} \right) = \frac{2(B-1)v^2}{2(1+B)v^3+g}  . 
 \label{eq:3-1.100}
\end{equation}
Therefore, the stability condition of the fixed point  is 
\begin{equation}
\left\{
\begin{array}{cc}
g>g^{*} &~~~ \text{for } B<1 \\
g<g^{*} &~~~ \text{for } B>1
\end{array}
\right. , 
 \label{eq:3-1.101}
\end{equation}
where the bifurcation threshold $g^{*}$ is determined as follows. 
From Eqs.~(\ref{eq:3-1.17}) and (\ref{eq:N1.3}), $v(>0)$ at the threshold is obtained as 
\begin{equation}
v = \left(\frac{\gamma-\kappa} {1-B} \right)^{1/2} , 
 \label{eq:N1.4}
\end{equation}
provided that  $(\gamma-\kappa)/(1-B)>0$. 
Substituting $v$ into Eq.~(\ref{eq:N1.3}), one obtains the bifurcation boundary as 
\begin{equation}
g^{*} \equiv \frac{2B (\gamma-\gamma_c)}{(1-B)} \left(\frac{\gamma-\kappa} {1-B} \right)^{1/2}  . 
 \label{eq:N1.5}
\end{equation}
Since $g^{*}>0$, the following condition should be satisfied;
\begin{equation}
\left\{
\begin{array}{cc}
B < 1 &~~~ \text{for } \gamma > \gamma_c \\
B > 1 &~~~ \text{for } \gamma < \gamma_c
\end{array}
\right.  . 
 \label{eq:N1.6}
\end{equation}  
The bifurcation threshold (\ref{eq:N1.5}) 
is indicated by the thick solid line in Figs.~\ref{fig:diagram_1}(a) and (b). Note that $B=1.25$ for $\kappa=0.2$ and $B=1/3$ for $\kappa=0.75$.  The stability condition (\ref{eq:3-1.101}) and the bifurcation lines agree with the numerical results obtained from the original equations  (\ref{eq:1.3}) - (\ref{eq:1.6}).

When $\gamma > \gamma_c$, there exists another bifurcation from the circular-drift motion to a zigzag-1 motion  as shown in  Figs.~\ref{fig:diagram_1}(a) and (b). 
We have obtained the boundary of this bifurcation by numerical simulations of the reduced equations (\ref{eq:3-1.1}) and (\ref{eq:3-1.2}). The results are plotted by the thin solid lines in Figs.~\ref{fig:diagram_1}(a) and (b). The line for $\kappa=0.75$ is in an apparent agreement with the numerical results of the original equations (\ref{eq:1.3}) - (\ref{eq:1.6}). See, however, further discussion  given in the next paragraph.
The thin solid line for $\kappa=0.2$ in Fig.~\ref{fig:diagram_1}(a) also agrees with the stability threshold of the circular-drift motion but the reduced equations do not reproduce the coexistence of the zigzag-1, zigzag-2 and circular-drift motions.

\begin{figure}[t]
  \begin{center}
  	\resizebox{0.45\textwidth}{!}{
	  \includegraphics{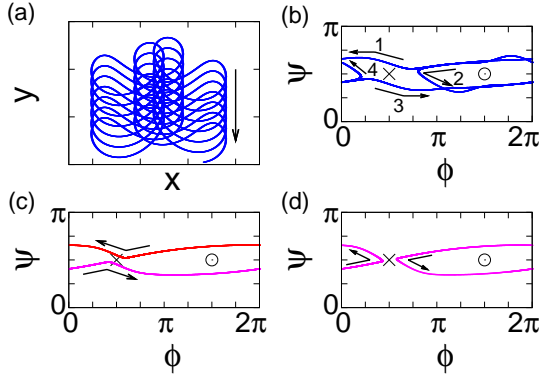}
	}
    \caption{(Colour on-line)
      Trajectory in the vicinity of the bifurcation between the circular-drift motion and the zigzag-1 motion for $\gamma = 3$ and  $\kappa = 0.75$ (a) in the real space and (b)-(d) in the $\phi$-$\psi$ space. 
      (a) and (b) are obtained from the original equations (\ref{eq:1.3}) - (\ref{eq:1.6}) for $g=0.232$ whereas  (c) and (d) are obtained from  Eqs.~(\ref{eq:3-1.1}) and (\ref{eq:3-1.2}) for $g=0.245132$ and $0.245133$ respectively. 
      The cross and circle in Fig.~(b) - (d) indicate the saddle point and the unstable fixed point, respectively. 
    } \label{fig:saddle_homoclinic}
  \end{center}
\end{figure}

We have examined the motion for $\kappa=0.75$ in the vicinity 
of the bifurcation between the circular-drift motion and the zigzag-1 motion.
The trajectory in the real space obtained numerically  from the original equations (\ref{eq:1.3}) - (\ref{eq:1.6}) for $\gamma=3$ and $g=0.232$ is displayed in Fig.~\ref{fig:saddle_homoclinic}(a). It is clear that this is neither a simple circular-drift motion nor a zigzag-1 motion but is a kind of a mixture of these two motions, or a mixture of a circular-drift motion and a zigzag-2 motion.  
In fact,  the trajectory in the $\phi$-$\psi$ space shown in  Fig.~\ref{fig:saddle_homoclinic}(b) is found to be a superposition of  the circular-drift motion and the zigzag-2 motion. See Figs.~\ref{fig:phipsi_1}(c) and (d).
The bifurcation for the reduced equations (\ref{eq:3-1.1}) and (\ref{eq:3-1.2}) is at about $g=0.245$ for $\kappa=0.75$ and $\gamma=3$. The trajectories near this threshold are shown in Fig.~\ref{fig:saddle_homoclinic} for (c) $g=0.245132$ and (d) $g=0.245133$. There are two trajectories in Fig.~\ref{fig:saddle_homoclinic}(c) corresponding to a right-moving and a left-moving circular-drift motions but these two merge each other near the saddle point in Fig.~\ref{fig:saddle_homoclinic}(d) 
to cause a zigzag-1 motion. Therefore this is a saddle homoclinic orbit bifurcation. The bifurcation exhibited in the original set of equations is more complicated.

\section{Analysis of the bifurcations II}\label{sec:analysis_bifurcations_2}

In this section, we analyze the bifurcations obtained by the original equations (\ref{eq:1.3}) - (\ref{eq:1.6}) with $g=0$. 
As in the preceding section, we consider a simplified set of equations eliminating $s$ and $v$.
From Eqs.~(\ref{eq:1.3}) - (\ref{eq:1.6}), by putting $d v /d t =d s / d t=0$, we obtain 
\begin{gather}
\frac{d \phi}{d t} =  F\left( \phi, \psi \right)
 \label{eq:3-2.1} \\
\frac{d \psi}{d t} =  G\left( \phi, \psi \right),  
 \label{eq:3-2.2}
\end{gather}
with 
\begin{gather}
F\left( \phi, \psi \right)  = 
 -\frac{as}{2}  \sin 2\psi 
 \label{eq:3-2.3}  \\
G\left( \phi, \psi \right) 
 =  -\frac{ b v^2 -a s^2}{2 s} \sin 2\psi -\frac{h}{2 s} \sin 2(\phi+\psi)  ,
 \label{eq:3-2.4}
\end{gather}
where $s$ is given by 
\begin{equation}
s(\phi,\psi) = \frac{b}{\kappa} v^2 \cos2\psi +\frac{h}{\kappa} \cos2(\phi+\psi)
 \label{eq:3-2.102}
\end{equation}
and $v$ is the largest positive root of a cubic equation 
\begin{equation}
(1+B\cos^22\psi)v^3 -\Gamma v = 0  . 
 \label{eq:3-2.103}
\end{equation}
We have defined 
\begin{equation}
\Gamma(\phi,\psi)\equiv\gamma -(h /b)B \cos 2(\phi+\psi) \cos 2\psi . 
 \label{eq:3-2.101}
\end{equation}
From Eqs.~(\ref{eq:3-2.103}) and (\ref{eq:3-2.102}),   $v$ and $s$ are given for  $\Gamma > 0$ by 
\begin{gather}
v\left( \phi, \psi \right) 
 = \left( \frac{ \Gamma(\phi,\psi) }{ 1+B\cos^2 2\psi } \right)^{1/2}
 \label{eq:3-2.5} \\
s\left( \phi, \psi \right) 
 = \frac{b \gamma \cos 2\psi +h \cos 2(\phi+\psi)}{ \kappa (1+B\cos^2 2\psi) }
 \label{eq:3-2.6}
\end{gather}
and for  $\Gamma \leq 0$ by
\begin{gather}
v\left( \phi, \psi \right) = 0 
 \label{eq:3-2.7} \\
s\left( \phi, \psi \right) = (h/\kappa) \cos 2(\phi+\psi)  . 
 \label{eq:3-2.8}
\end{gather} 
In both cases, $s\ge0$ is required. 
By the numerical simulations of the reduced equations (\ref{eq:3-2.1}) and (\ref{eq:3-2.2}),  all types of motions except for the zigzag-2 motion are obtained.

As in section  \ref{sec:analysis_bifurcations_1}, we have verified numerically that  the zigzag-2 motion is realized if we retain $s$ as a slow variable;
\begin{gather}
\frac{d \phi}{d t} = -\frac{a}{2} s \sin 2\psi 
 \label{eq:3-2.31} \\
\frac{d s}{d t} = -\kappa s +b v^2 \cos 2\psi + h \cos 2(\phi+\psi)
 \label{eq:3-2.32} \\
\frac{d \psi}{d t} = -\frac{bv^2-as^2}{2 s}  \sin 2\psi -\frac{h}{2 s} \sin 2(\phi+\psi)
 \label{eq:3-2.33} 
\end{gather}
where
\begin{equation}
v = 
\left\{
\begin{array}{cr}
(\gamma-\frac{a}{2}s \cos 2\psi)^{1/2}~ & \text{if } \gamma-\frac{a}{2}s \cos 2\psi\ge0 \\
0 & \text{if } \gamma-\frac{a}{2}s \cos 2\psi<0
\end{array}
\right.
 \label{eq:3-2.34}
\end{equation}
However, in the theoretical analysis given below, we employ Eqs.~(\ref{eq:3-2.1}) and (\ref{eq:3-2.2}) which are much easier to treat.

In the restricted space $0<\phi<2\pi$ and $0<\psi<\pi$, 
Eqs.~(\ref{eq:3-2.1}) and (\ref{eq:3-2.2}) have four fixed points,  $(\phi,\psi) = (n\pi/2,\bar{\psi}_0)$ with $n=0,1,2,3$.
In order to satisfy the condition  $s>0$, we note from Eqs.~(\ref{eq:3-2.6}) and (\ref{eq:3-2.8}) that $\bar{\psi}_0=\pi/2$ for $n=1$ and 3 while for $n=0$ and 2, $\bar{\psi}_0=0$ when $b\gamma+h>0$ and $\bar{\psi}_0=\pi/2$ when $b\gamma+h<0$. 
If $\Gamma=b\gamma-hB>0$, i.e. $v>0$, the two solutions  $(\phi,\psi) = (n\pi/2,\bar{\psi}_0)$ with $n=0$ and 2 represent a straight motion propagating to the right and the left, respectively, 
 whereas, if $\Gamma\le0$, i.e. $v=0$, they 
 are a pair of degenerate solutions of the motionless state.

We introduce the linear stability matrix in the same form as Eq.~(\ref{eq:3-1.10})
and analyze the stability of the four  fixed points. 
In the region $\Gamma > 0$, $v$ and $s$ are given by Eqs.~(\ref{eq:3-2.5}) and (\ref{eq:3-2.6}) respectively. 
From the stability matrix around the fixed points  $(0, \bar{\psi_0})$ and $(\pi, \bar{\psi}_0)$, 
 the trace and the determinant of $L( \phi, \psi )$ are given, respectively, by 
\begin{gather}
\tr L_{1+} = \frac{2B (\gamma-\gamma_c) +a h /\kappa } {(1+B)} 
 \label{eq:3-2.10} \\
\det L_{1+} = - a h  . 
 \label{eq:3-2.11}
\end{gather}
When $a<0$ and $h>0$ as we have assumed, the trace is positive if $0<h<h_{H}\equiv |b| (\gamma-\gamma_c)$ for $\gamma>\gamma_c$ and negative if $h>h_{H}$ for $\gamma>\gamma_c$ or for $\gamma<\gamma_c$, whereas the determinant is always positive. 
Therefore, the fixed points $(0, \bar{\psi}_0)$ and $(\pi, \bar{\psi}_0)$ corresponding to a straight motion are always stable if $\gamma<\gamma_c$ as expected. 
When $\gamma>\gamma_c$, they are stable for $h>h_{H}$ and they lose their stability for $0<h<h_{H}$ by a Hopf bifurcation. This corresponds to a transition from a straight motion for $h>h_{H}$ to a zigzag-1 motion for  $0<h<h_{H}$. See Figs.~\ref{fig:phipsi_2}(a) and (b).
The bifurcation boundary $h_{H}$ represented by the thick solid line in Fig.~\ref{fig:diagram_2}, is found to agree with the results of the numerical simulations of the original equations (\ref{eq:1.3}) - (\ref{eq:1.6}).

There is another bifurcation in the region where the fixed points $(0, \bar{\psi}_0)$ and $(\pi, \bar{\psi}_0)$ for the straight motion are stable.
The requirement $s\ge0$ leads  from Eq.~(\ref{eq:3-2.6}) to 
\begin{equation}
\bar{\psi}_0 = 
\left\{
\begin{array}{ccl}
\pi/2 &~~~~~~~~ \text{for } & h<h^{*}\equiv |b|\gamma \\
0 &~~~~~~~~ \text{for } & h>h^{*}
\end{array}
\right.
 \label{eq:3-2.100}
\end{equation}
This means that  elongation of the particle is  perpendicular (parallel) to the velocity (external force) for $h<(>) h^{*}$.
Since we have assumed $h>0$, this occurs only for $\gamma>0$. 
This bifurcation threshold  represented by the thick dotted line in Fig.~\ref{fig:diagram_2} is consistent with the results of  numerical simulations of the original equations (\ref{eq:1.3}) - (\ref{eq:1.6}).

The determinant and the trace of the stability matrix of the other two fixed points $(\pi/2, \bar{\psi}_0)$ and $(3\pi/2,  \bar{\psi}_0)$ are given, respectively, by 
\begin{gather}
\tr L_{2+} = \frac{2B (\gamma-\gamma_c) -a h /\kappa } {(1+B)} 
 \label{eq:3-2.13} \\
\det L_{2+} = a h . 
 \label{eq:3-2.14}
\end{gather}
The determinant is always negative for $a<0$ and $h>0$. This implies that the fixed points $(\pi/2, \bar{\psi}_0)$ and $(3\pi/2,  \bar{\psi}_0)$ are saddle points.

In the region $\Gamma < 0$, 
we can show by a similar analysis that the fixed-points $(0, \bar{\psi}_0)$ and $(\pi, \bar{\psi}_0)$ with $\bar{\psi}_0=0$ are stable, whereas the other two fixed-points $(\pi/2,\bar{\psi_0})$ and $(3\pi/2,\bar{\psi}_0)$ are saddle.

\begin{figure}[t]
  \begin{center}
	\resizebox{0.45\textwidth}{!}{
	  \includegraphics{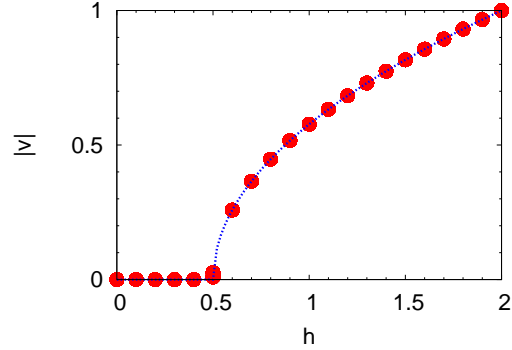}
	}
    \caption{(Colour on-line)
      Migration velocity in the vicinity of  the pitchfork bifurcation boundary $\gamma = -0.5$. 
      The dots indicate the results obtained by solving Eqs.~(\ref{eq:1.3}) - (\ref{eq:1.6}) numerically and the dotted line indicates the velocity obtained analytically from Eq.~(\ref{eq:3-2.20}). 
    } \label{fig:pitchfork}
  \end{center}
\end{figure}

The bifurcation from the motionless state to the straight motion in Fig.~\ref{fig:diagram_2}  is derived as follows.
For the stable straight motion  $(\phi, \psi)=(0,0)$ and $(\pi,0)$, $\Gamma$ is given by $\Gamma=\gamma -(h/b)B$. We note from Eq.~(\ref{eq:3-2.103}) that a bifurcation from the state $v=0$ to $v\ne0$ occurs at $\gamma =(h/b)B$.
Since  $h>0$, $B>0$ and $b<0$, this bifurcation is possible only for $\gamma<0$. 
Therefore, when $h$ is increased, the bifurcation occurs at $h=h_p$, where 
\begin{equation}
h_p = -\frac{|b| \gamma }{B}   . 
 \label{eq:3-2.19}
\end{equation}
This bifurcation boundary indicated by the thin solid line in Fig.~\ref{fig:diagram_2} agrees quite well  with the results of  numerical simulations of the original equations (\ref{eq:1.3}) - (\ref{eq:1.6}). 
The velocity around the pitchfork bifurcation is given from Eq.~(\ref{eq:3-2.5}) by 
\begin{equation}
v_{pf} =\left( \frac{\gamma+h B /|b|} {1+B} \right)^{1/2}  . 
 \label{eq:3-2.20}
\end{equation}
This theoretical result is plotted in Fig.~\ref{fig:pitchfork} in comparison with   the numerical result of Eqs.~(\ref{eq:1.3}) - (\ref{eq:1.6}).

Finally, we mention that the thin dotted line in Fig.~\ref{fig:diagram_2} is the boundary between the circular motion and the zigzag-1 motion obtained numerically from Eqs.~(\ref{eq:3-2.1}) and (\ref{eq:3-2.2}). Since this set of reduced equations does not reproduce the zigzag-2 motion, the agreement with the results of the original equations (\ref{eq:1.3}) - (\ref{eq:1.6}) is poor.

To summarize, the reduced equations  (\ref{eq:3-2.1}) and (\ref{eq:3-2.2}) provide very accurately the three bifurcation thresholds
given by the thick solid line, the thick dotted line and the thin solid line in Fig.~\ref{fig:diagram_2}. However, in order to reproduce theoretically the zigzag-2 motion, we have to take into consideration of  the variable $s$.

\section{Discussion}\label{sec:discussion}

We have  investigated dynamics of a self-propelled particle under two types of external forcing, the gravitational-like force and the electric-like force. 
In both cases, we have found a variety of dynamical motions and have obtained numerically the dynamical phase diagrams (Figs.~\ref{fig:diagram_1} and \ref{fig:diagram_2}). 
We have also analyzed the bifurcations of these motions by using the reduced equations in terms of the two kinds of angles which represent the migration velocity and the elongation of a particle.

In the case of the gravitational-like force, we have found the circular-drift motion, the zigzag-1 motion, the zigzag-2 motion, and the straight-falling motion. 
The reduced equations (\ref{eq:3-1.1}) and (\ref{eq:3-1.2})  admit these solutions except for  the zigzag-2 motion.
We have shown that a Hopf bifurcation appears  between the zigzag-1 motion and the straight-falling motion. 
This property agrees  with the numerical results of the original equations (\ref{eq:1.1}) and (\ref{eq:1.2}) with $Q_{\alpha\beta}=0$ as shown in Fig.~\ref{fig:diagram_1}.

In the case of the electric-like force, we have obtained the circular motion, the zigzag-1 motion, the zigzag-2 motion, and two types of the straight motion, whose direction of deformation  is either parallel or perpendicular to the velocity vector.
The reduced equations (\ref{eq:3-2.1}) and (\ref{eq:3-2.2}) reproduce all of these motions except for the zigzag-2 motion.  It has been shown that there appears a Hopf bifurcation  between the zigzag-1 motion and the straight motion, and a pitchfork bifurcation between the motionless state and the straight motion. 
These properties are consistent with the numerical results of the original equations (\ref{eq:1.1}) and (\ref{eq:1.2}) with $g=0$ as shown in Fig.~\ref{fig:diagram_2}. In the both cases, however, the variable $s$ has to be considered as a slow variable to realize theoretically the zigzag-2 motion.

The basic ingredient of the systems studied is the competition between the external forcing and the circular motion which occurs in the absence of the external forces. This causes the transitions between the straight motion and the two types of zigzag motions and the dynamics of circular-drift motion. At present, we have no experiments which are directly related to the predictions.  However, since our time-evolution equations are general based on the symmetry argument, we expect that the predictions made in the present paper will be detected experimentally in the near future.

We are grateful to Professor H. Loewen for valuable discussions at the early stage of this work.
This work was supported by the JSPS Core-to-Core Program gInternational research network for non-equilibrium dynamics of soft matterh
and  by the Grant-in-Aid for the 21st Century
COE gCenter for Diversity and Universality in Physicsh and
the Grant-in-Aid for the superior area gSoft Matter Physicsh
both from the Ministry of Education, Culture, Sports, Science
and Technology (MEXT) of Japan.

\end{document}